\documentclass[12pt]{amsart}

\usepackage{graphicx}
\usepackage{psfrag}   
\usepackage{mathptmx} 
\usepackage{amssymb}
\usepackage{amsmath}
\usepackage{amsthm}
\usepackage{mathtools}
\usepackage[dvipsnames]{xcolor}
\usepackage[sort,round]{natbib}
\usepackage{marvosym}
\usepackage{a4wide}
\usepackage[normalem]{ulem}
\newcommand{\stkout}[1]{\ifmmode\text{\sout{\ensuremath{#1}}}\else\sout{#1}\fi}
\usepackage{enumerate}
\usepackage{url}
\usepackage{csquotes}

\usepackage[mathscr]{euscript}

\newtheorem{theorem}{Theorem}[section]
\newtheorem{lemma}[theorem]{Lemma} 
\newtheorem{corollary}[theorem]{Corollary}

\newtheorem{remark}[theorem]{Remark}

\newcommand{\cC}{{\mathcal C}}
\newcommand{\cS}{{\mathcal S}}

\newcommand{\pf}{\noindent{\em Proof: }}
\newcommand{\epf}{\hfill\hbox{\rule{3pt}{6pt}}\\}

\title{Phylogenetic diversity indices from an affine and projective viewpoint}
\author{V. Moulton}
\thanks{School of Computing Sciences, University of East Anglia, UK}
\author{A. Spillner}
\thanks{Merseburg University of Applied Sciences, Germany}
\author{K. Wicke}
\thanks{Department of Mathematical Sciences, New Jersey Institute of Technology, USA}

\date{\today}

\begin{document}

\begin{abstract}
Phylogenetic diversity indices are commonly used to rank the elements
in a collection of species or populations for conservation purposes.
The derivation of these indices is typically based on some
quantitative description of the evolutionary history of the
species in question, which is often given in terms of a phylogenetic tree. 
Both rooted and unrooted phylogenetic trees can be employed, and there are close
connections between the indices that are derived in these two different ways. 
In this paper, we introduce more general phylogenetic diversity indices that 
can be derived from collections of subsets (clusters) and
collections of bipartitions (splits) of the
given set of species. Such indices could be useful, for example, in case 
there is some uncertainty in the topology of 
the tree being used to derive a phylogenetic diversity index.
As well as characterizing some of the
indices that we introduce in terms of their special properties, we 
provide a link between
cluster-based and split-based phylogenetic diversity indices  
that uses
a discrete analogue of the classical link between affine and projective 
geometry. This provides a unified framework for
many of the various phylogenetic diversity indices used in the literature
based on rooted and unrooted phylogenetic trees, 
generalizations and new proofs for previous results
concerning tree-based indices, and a way to define some 
new phylogenetic diversity indices that naturally arise as 
affine or projective variants of each other.
\end{abstract}

\maketitle

\section{Introduction}
Evolutionary isolation metrics or phylogenetic diversity indices
provide quantitative measures of biodiversity and are increasingly
popular tools to prioritize species for conservation
\citep{isa-tur-07,red-har-08, red-maz-14, red-moo-06a, tuc-cad-16, van-wri-91}. 
These indices quantify the importance of a species to overall 
biodiversity by assessing its unique and shared evolutionary 
history as indicated by its placement in an underlying phylogeny. 
Preserving phylogenetic diversity and the \enquote{Tree of Life} has become an integral component of conservation considerations (see, e.g., the \enquote{Phylogenetic Diversity Task Force}\footnote{\url{https://www.pdtf.org/}} initiated by the IUCN).
Indeed, conservation initiatives like the EDGE
of Existence programme\footnote{\url{https://www.edgeofexistence.org/}}
\citep{gum-gra-23,isa-tur-07} incorporate phylogenetic diversity indices in their identification of species that are both evolutionary distinct and globally endangered.  
 Moreover, the \enquote{guide to phylogenetic metrics for conservation, community ecology and macroecology} by \citet{tuc-cad-16} has been cited more than 700 times since its publication, 
 thus demonstrating an even more widespread interest and application of phylogenetic tools, and
 in particular different phylogenetic diversity indices, within conservation settings.

Mathematically, with a multitude of phylogenetic diversity indices at hand,
there is now an increasing interest in understanding how the different
indices relate to each other. Much of the previous work in this
direction has focused on
comparing and analyzing different indices derived from rooted phylogenetic
trees \citep{bor-sem-23a, man-22a, man-ste-23a, wic-ste-19a}.
Phylogenetic diversity indices have also been defined for unrooted trees
\citep{haa-kas-08a,wic-ste-19a}, and an exploration of the relationship between
indices derived via rooted and unrooted phylogenetic trees
is presented by~\citet{wic-ste-19a}.

\begin{figure}
\centering
\includegraphics[scale=1.0]{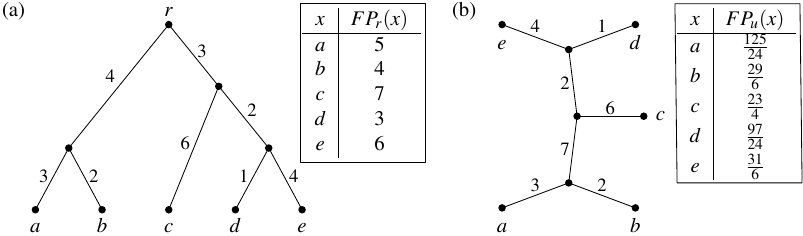}
\caption{(a)~A rooted phylogenetic tree on the set
\(X = \{a,b,c,d,e\}\) of species. The root vertex is~\(r\) and all
edges are weighted. The table gives the value \(FP_r(x)\) of the
fair proportion index on this rooted tree for each \(x \in X\).
(b)~The unrooted phylogenetic tree with weighted edges
on the same set \(X\) of species obtained by suppressing
the root of the tree in (a). The table gives the value~\(FP_u(x)\) of the
fair proportion index on this unrooted tree for each~\(x \in X\).}
\label{fig:fp:ex:rooted:unrooted}
\end{figure}

As one might expect, phylogenetic diversity indices for
rooted and unrooted trees are closely related.
To illustrate this, consider the much studied 
\emph{fair proportion index} \citep{isa-tur-07,red-03}.
For the rooted phylogenetic tree with edge weights in
Figure~\ref{fig:fp:ex:rooted:unrooted}(a),
the value \(FP_r(x)\) of the rooted
fair proportion index for a species \(x \in X\) is computed
by adding, over all edges that are contained
in the path from the root \(r\) to the leaf labeled by~\(x\),
the weight of the edge divided by the total
number of species for which the path from the root to the leaf
labeled by that species also contains that edge. For example,
for species \(e\) there are three edges in the path from \(r\) to \(e\)
and we obtain
\begin{equation}\label{eq:fpeg}
FP_r(e) = \frac{3}{3} + \frac{2}{2} + \frac{4}{1} = 6.
\end{equation}

In~\citep{wic-ste-19a} the fair proportion index has also
been defined for unrooted phylogenetic trees.
Consider the unrooted phylogenetic tree with edge weights in
Figure~\ref{fig:fp:ex:rooted:unrooted}(b).
The removal of an edge breaks the tree into two subtrees. The value
\(FP_u(x)\) of the unrooted fair proportion index
for a species \(x \in X\) is one half
of the value obtained by adding, over all edges in the unrooted tree,
the weight of the edge divided by the number of species
that lie in the same subtree as \(x\) after removal of the edge.
For example, for species \(a\) we obtain
\begin{equation}\label{eq:sfpeg}
FP_u(a) = \frac{1}{2} \cdot \left ( \frac{3}{1} + \frac{2}{4}
+ \frac{7}{2} + \frac{6}{4}
+ \frac{2}{3} + \frac{1}{4} + \frac{4}{4} \right ) = \frac{125}{24}.
\end{equation}

While, at first glance, there is only some
similarity in the way the values of the fair
proportion index for rooted and unrooted 
phylogenetic trees are computed from the edge weights, there is a
deeper connection. For example, as can be seen
in Figure~\ref{fig:fp:ex:rooted:unrooted},
\(\sum_{x \in X} FP_r(x) = \sum_{x \in X} FP_u(x) = 25\),
which is the total weight of the edges of the phylogenetic tree
from which the values are computed. 

To better understand this type of observation, in this paper we
consider indices from the viewpoint of affine and projective
clustering. This way of thinking about clustering has its origins 
in~\citep{dre-97a}, and since then has become a useful tool
in phylogenetic combinatorics
(see, e.g., \citealt[Ch.~9]{dress2012basic}
and \citealt{kleinman2013affine}).
More specifically, in this paper we extend the study of
phylogenetic diversity indices into the more   
general setting of collections of \emph{clusters} (subsets of a set) and 
collections of \emph{splits} (bipartitions of a set). These 
settings correspond to affine and projective viewpoints 
of clustering, respectively (see Section~\ref{sec:framework}).
Considering collections of clusters and
splits in general can be beneficial since it
allows for the representation of data that is not tree-like or
where it is difficult to determine the correct topology for a 
phylogenetic tree. Indeed, phylogenetic diversity indices
have already been introduced for split systems
(see, e.g., \citealt{abh-col-23a}).

To illustrate this way of 
thinking, as hinted above, collections of clusters naturally 
arise when computing the rooted fair proportion index.
In particular, clusters arise from rooted 
phylogenetic trees by taking, for each edge,
the subset of species for which the path from the root to
that species contains the edge
(e.g. in  in Figure~\ref{fig:fp:ex:rooted:unrooted}(a) 
the edge with weight~3 next to the root
gives rise to the cluster $\{c,d,e\}$). 
Thus, the sum used to compute the fair proportion
index of $e$ in Equation~(\ref{eq:fpeg}) is just the sum
of the values $\frac{\omega(C)}{|C|}$ taken over
all clusters $C$ that contain $e$, where~$\omega(C)$ is
the weight of the edge giving rise to cluster $C$ and
\(|C|\) denotes the number of species in~\(C\).
Similarly, we can interpret Equation~(\ref{eq:sfpeg}) in terms of splits, 
using the fact that splits arise from unrooted phylogenetic trees by taking,
for each edge, the split obtained by removing the edge
and considering the subsets of species in the two resulting subtrees
(e.g. in Figure~\ref{fig:fp:ex:rooted:unrooted}(b) the
edge with weight~7 gives rise to
the split $\{\{a,b\},\{c,d,e\}\}$). Then the
sum used to compute the unrooted fair proportion
index of $a$ in Equation~(\ref{eq:sfpeg}) is just the sum
of the values $\frac{\lambda(S)}{2|A|}$ taken over all  
splits \(S\) coming from the tree, where \(\lambda(S)\)
is the weight of the edge giving rise to~\(S\) and~\(A\)
is the part in $S$ that contains~$a$.
Clearly, the sums used to compute $FP_r$ and 
$FP_u$ can be generalized to any collection of weighted
clusters or splits, respectively,
and, as we shall see, the more general phylogenetic diversity indices that 
result in this way have similar properties to their 
tree-based counterparts.

Thinking about phylogenetic diversity
indices in an affine and projective way, leads us to two key
questions that we will consider in this paper:
\begin{itemize}
\item[(i)] How do properties of tree-based phylogenetic diversity
indices extend to indices defined via collections of clusters and splits?
\item[(ii)] How can the relationships between collections 
of clusters and collections of splits be exploited to
relate cluster- and split-based phylogenetic diversity indices?
\end{itemize}
In this contribution, we give answers to both of
these questions, introducing the 
concept of phylogenetic diversity indices based on
collections of clusters and splits, and giving characterizations 
for some of these indices in terms of their special properties. We also 
present a general framework to systematically relate
cluster- and split-based phylogenetic diversity indices
via a process that is commonly used in phylogenetic combinatorics.
This provides concise proofs for generalizations of
previous results for trees as well as ways to define new indices.
We illustrate our new concepts and results by focusing
on a few well-known tree-based phylogenetic diversity
indices including the fair proportion index,
the Shapely value \citep{haa-kas-08a,sha-53a}, and
the equal splits index \citep{red-moo-06a}. 

The rest of this paper is structured as follows.
In Section~\ref{sec:clusters} we formally define 
cluster-based phylogenetic diversity indices and
present some key properties that such indices may have.
Then, in Section~\ref{sec:FPcharacterization}, we
present a characterization of the general
cluster-based fair proportion index. In
Section~\ref{sec:shapely} we present a
characterization of the Shapley value and
use its relationship to the fair proportion index
to describe the first building block of our framework.
In Section~\ref{sec:framework} we then give the
complete framework, and illustrate some of its
applications in Section~\ref{sec:equal:splits}
using the fair proportion index and the 
equal splits index as examples. We conclude in
Section~\ref{sec:conclusion} discussing some potential interesting 
directions for future work.
	
\section{Cluster-based indices}
\label{sec:clusters}

Let \(X\) be a non-empty finite set. We denote the
power set of \(X\) by \(\mathcal{P}(X)\). We call a 
non-empty subset \(C \subseteq X\) a \emph{cluster} on \(X\)
and call a non-empty collection
\(\mathcal{C} \subseteq \mathcal{P}(X) \setminus \{\emptyset\}\)
a \emph{cluster system} on~\(X\). In this
section we introduce the concept of a phylogenetic diversity
index on a cluster system, and illustrate some basic 
properties of these indices by considering a generalization of the
fair proportion index for rooted trees that we introduced in the
introduction.

To motivate the definition of a phylogenetic diversity
index on a cluster system, we briefly look again
at rooted phylogenetic trees. 
Fixing a rooted phylogenetic tree \(\mathcal{T}\)
on a set \(X\) of species, a phylogenetic diversity
index \(\Phi\) on \(\mathcal{T}\) assigns, to each weighting \(\omega\)
of the edges in \(\mathcal{T}\), a vector 
\(\Phi(\omega) \in \mathbb{R}^X\). To give an example,
let \(\Phi\) be the fair proportion index on the rooted
phylogenetic tree in Figure~\ref{fig:fp:ex:rooted:unrooted}(a).
Then, for the weighting~\(\omega\) of its edges given in
Figure~\ref{fig:fp:ex:rooted:unrooted}(a), we can write
\begin{equation}
\label{eq:vector:fp:example:intro}
\Phi(\omega) = (5,4,7,3,6),
\end{equation}
or, in more detail,
\((\Phi(\omega))(a) = 5, \ (\Phi(\omega))(b) = 4,\dots,(\Phi(\omega))(e) = 6\).

\begin{figure}
\centering
\includegraphics[scale=1.0]{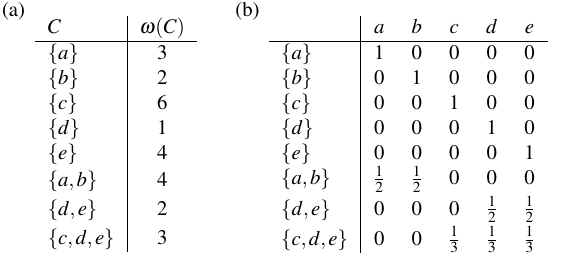}
\caption{(a) The weighted clusters on \(X\) corresponding
to the edges of the rooted phylogenetic tree in
Figure~\ref{fig:fp:ex:rooted:unrooted}(a).
(b) The matrix~\(\Gamma\) from Equation~(\ref{eq:matrix:linear:pdi})
for the fair proportion index on \(\mathcal{C}\),
where \(\mathcal{C}\) is the cluster system consisting
of the clusters given in~(a).}
\label{fig:gamma:fp:ex}
\end{figure}

As described in the introduction,
each edge in a rooted phylogenetic tree on~\(X\)
is associated with a cluster on~$X$.  
In Figure~\ref{fig:gamma:fp:ex}(a) the clusters
associated with the edges of the rooted phylogenetic tree in
Figure~\ref{fig:fp:ex:rooted:unrooted}(a) are given, where
each cluster is weighted by the length of the corresponding edge.
Note that this cluster system~\(\mathcal{C}\) has a special property, namely 
it is a \emph{hierarchy}, that is, 
\(C \cap C' \in \{\emptyset,C,C'\}\) holds for
all \(C,C' \in \mathcal{C}\). 
In particular, as we see in this example, hierarchies are
essentially those cluster systems that can be represented by a
rooted phylogenetic tree on~\(X\)
(see, e.g., \citealt[Thm.~3.5.2]{SS03} for a more precise statement
of this fact using the concept of a rooted \(X\)-tree).

Bearing these facts in mind, 
for an arbitrary cluster system \(\mathcal{C}\) on \(X\),
we consider the space~$\mathbb{L}(\mathcal{C})$ consisting
of all weightings $\omega: \mathcal{C} \rightarrow \mathbb{R}$. 
We then define a \emph{phylogenetic diversity index} on~\(\mathcal{C}\) to be a 
map \(\Phi: \mathbb{L}(\mathcal{C}) \rightarrow \mathbb{R}^X\).
For example, following the intuitive description in
the introduction, we define the \emph{fair-proportion index} on
a cluster system~\(\mathcal{C}\) on~$X$ by putting,
for each \(\omega \in \mathbb{L}(\mathcal{C})\) and
all \(x \in X\),
\begin{align}
(FP(\omega))(x) &= \sum\limits_{C \in \mathcal{C}: \ x \in C} \frac{\omega(C)}{\vert C \vert}.
\label{def:fp}
\end{align}
It can then be checked that~(\ref{def:fp}) applied to
the weighted cluster system in Figure~\ref{fig:gamma:fp:ex}(a)
yields precisely the vector we saw in~(\ref{eq:vector:fp:example:intro}).

We now introduce three key properties of cluster-based indices
which generalize properties of tree-based indices described in the
literature. We will illustrate these properties for the 
fair proportion index and, as we shall see, 
these properties are also shared
by some of the other phylogenetic diversity indices
that we consider later on.

Let \(\mathcal{C}\) be a cluster system on~\(X\).
A phylogenetic diversity index~\(\Phi\) on~\(\mathcal{C}\)
is \emph{additive} if
\begin{itemize} 
\item[(A)]
\(\Phi(\omega_1 + \omega_2) = \Phi(\omega_1) + \Phi(\omega_2)\)
for all \(\omega_1, \omega_2 \in \mathbb{L}(\mathcal{C})\),
\end{itemize}
and \(\Phi\) is \emph{homogeneous} if
\begin{itemize}
\item[(H)]
\(\Phi(a \cdot \omega) = a \cdot \Phi(\omega)\)
for all \(\omega \in \mathbb{L}(\mathcal{C})\) and all \(a \in \mathbb{R}\).
\end{itemize}
Properties~(A) and~(H) together mean that \(\Phi\) is
a \emph{linear map}, in which case we call \(\Phi\) \emph{linear}.
Linearity is a natural property to require
even on a rooted phylogenetic
tree~\(\mathcal{T}\), where it implies, for example, that,
applying the phylogenetic diversity index to a weighting
obtained by taking the average over several different edge
weightings of~\(\mathcal{T}\) amounts to averaging
the phylogenetic diversity index.
Note that every linear phylogenetic diversity index~\(\Phi\) on 
\(\mathcal{C}\) corresponds to
a \(|\mathcal{C}| \times |X|\)-matrix
\(\Gamma = \Gamma_{\Phi} = (\gamma_{(C,x)})\) such
that
\begin{equation}
\label{eq:matrix:linear:pdi}
(\Phi(\omega))(x) = \sum_{C \in \mathcal{C}} \omega(C) \cdot \gamma_{(C,x)}
\end{equation}
for all \(\omega \in \mathbb{L}(\mathcal{C})\) and all \(x \in X\).
For example, the matrix~\(\Gamma\) corresponding to
the fair proportion index on the cluster system in
Figure~\ref{fig:gamma:fp:ex}(a) is given in
Figure~\ref{fig:gamma:fp:ex}(b).
Finally, we call a phylogenetic diversity index~\(\Phi\)
on~\(\mathcal{C}\) \emph{complete} if
\begin{itemize}
\item[(C)]
\(\sum_{x \in X} (\Phi(\omega))(x) = \sum_{C \in \mathcal{C}} \omega(C)\)
holds for all \(\omega \in \mathbb{L}(\mathcal{C})\).
\end{itemize}
For tree-based phylogenetic diversity indices,
completeness is often required as part
of their definition (see, e.g., \citealt{bor-sem-23a,wic-ste-19a}).
For example, we have seen in the introduction for the
fair proportion index on a rooted phylogenetic tree
that \(\sum_{x \in X} FP_r(x)\) equals the total weight
of the edges in the tree. Property~(C) expresses this
fact in terms of clusters.
Note that if a linear phylogenetic diversity index \(\Phi\)
is complete, then we have \(\sum_{x \in X} \gamma_{(C,x)} = 1\)
for all \(C \in \mathcal{C}\) (cf. \citealt[Eq. (2)]{wic-ste-19a}), 
where \(\Gamma = (\gamma_{(C,x)})\)
is the matrix from Equation~(\ref{eq:matrix:linear:pdi}).
 
We conclude this section by showing that the
fair proportion index satisfies all three of the above properties.

\begin{lemma}
\label{lem:FPproperties}
The fair proportion index is a complete, linear
phylogenetic diversity index on \(\mathcal{C}\)
for any cluster system~\(\mathcal{C}\) on~\(X\).
\end{lemma}

\pf
As we have seen in the example in Figure~\ref{fig:gamma:fp:ex},
the fair proportion index can be described by a matrix
\(\Gamma = (\gamma_{(C,x)})\) where the row associated with
a cluster \(C \in \mathcal{C}\) contains \(|C|\) entries
equal to~\(\frac{1}{|C|}\) and \(|X| - |C|\) entries
equal to~0. 
\epf

\section{A characterization of the fair proportion index}
\label{sec:FPcharacterization}

In general, it is of interest to characterize
phylogenetic diversity indices in terms of
their key properties, as this can help
to understand better how they are related to one another. 
In this section, as an illustration for cluster-based indices, 
we shall present a characterization of the fair proportion index.
This generalizes the characterization of
the fair proportion index on rooted phylogenetic trees
given by~\citet[Thm.~6]{man-ste-23a}.  

Our characterization will require three properties.
The first two properties concern linear phylogenetic 
diversity indices \(\Phi\) on a cluster system \(\mathcal{C}\)
on \(X\), and are given in terms 
of the matrix corresponding to \(\Phi\).
For all \(C \in \mathcal{C}\), let \(ch(C)\) denote
the set of those \(C' \in \mathcal{C}\) with \(C' \subsetneq C\)
such that there is no \(C'' \in \mathcal{C}\) with
\(C' \subsetneq C'' \subsetneq C\).
We say that~\(\Phi\) satisfies the \emph{neutrality condition} if
\begin{itemize}
\item[(NC)]
the entries of the matrix~\(\Gamma_{\Phi}\) in
Equation~(\ref{eq:matrix:linear:pdi}) are such that
$\gamma_{(C,x)} = \gamma_{(C,y)}$
holds for all $C \in \mathcal{C}$ with $ch(C) = \emptyset$
and all $x,y \in C$. 
\end{itemize}
A property similar to (NC) was introduced by~\citet{man-ste-23a}
for rooted \(X\)-trees. In addition, we say that~\(\Phi\)
is a \emph{descendant diversity index} if
\begin{itemize}
\item[(DD)]
\(\Phi\) is complete,
all entries of the matrix~\(\Gamma_{\Phi}\)
in Equation~(\ref{eq:matrix:linear:pdi})
are non-negative and, for all \(C \in \mathcal{C}\),
\(\gamma_{(C,x)} = 0\) if \(x \not \in C\).
\end{itemize}
Property~(DD) was introduced by \citet{bor-sem-23a}
for the special case where the cluster system~\(\mathcal{C}\)
is a hierarchy (using the equivalent description of hierarchies
in terms rooted \(X\)-trees). 

\begin{figure}
\centering
\includegraphics[scale=1.0]{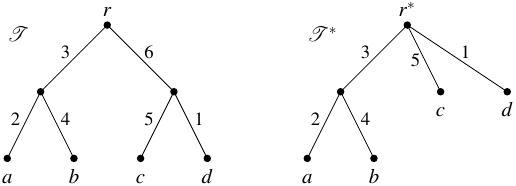}
\caption{Collapsing the edge with weight~\(6\) in the
rooted phylogenetic tree~\(\mathcal{T}\) on \(X=\{a,b,c,d\}\)
yields the rooted phylogenetic tree~\(\mathcal{T}^*\) on~\(X\).}
\label{fig:collapse:edge:ex}
\end{figure}

The third property is a bit more complicated, and 
thus we first motivate it using rooted trees as
in~\citep{man-ste-23a}.
Let \(\mathcal{T}\) be a rooted phylogenetic tree on~\(X\)
with edge weights and let \(\mathcal{T}^*\) be the
rooted phylogenetic tree on~\(X\) obtained by collapsing
one of the edges of \(\mathcal{T}\). This is illustrated
in Figure~\ref{fig:collapse:edge:ex}. In addition, let
\(\Phi\) and \(\Phi^*\) be phylogenetic diversity indices
on \(\mathcal{T}\) and \(\mathcal{T}^*\), respectively.
Both \(\Phi\) and \(\Phi^*\) yield a vector in \(\mathbb{R}^X\) for all
weightings of the edges of~\(\mathcal{T}\) and~\(\mathcal{T}^*\),
respectively. The topology of the rooted phylogenetic trees,
however, may have an impact on how the weights of the edges are used to
compute these vectors by~\(\Phi\) and~\(\Phi^*\),
respectively. Therefore, since the topologies of
\(\mathcal{T}\) and \(\mathcal{T}^*\) differ,
the vector in \(\mathbb{R}^X\) that we obtain by
\(\Phi^*\) for \(\mathcal{T}^*\) will usually not coincide
with the vector that we obtain in the limit,
as the weight of the edge in~\(\mathcal{T}\)
tends to~0, by~\(\Phi\) (keeping the weights
of all other edges in~\(\mathcal{T}\) in constant).

With this in mind, let~\(\mathcal{C}\) be a cluster
system on~\(X\) and let \(C \in \mathcal{C}\) such that
\(\mathcal{C}^* = \mathcal{C} \setminus \{C\}\) is non-empty.
A phylogenetic diversity index \(\Phi\) on \(\mathcal{C}\)
is \emph{downward continuous} with respect to a phylogenetic
diversity index \(\Phi^*\) on~\(\mathcal{C}^*\) if
\begin{itemize}
\item[(DC)]
for all \(\omega \in \mathbb{L}(\mathcal{C})\) we have
\begin{equation}
\label{eq:limit:continuous}
\lim_{\omega(C) \rightarrow 0} \Phi(\omega) = \Phi^*(\omega^*),
\end{equation}
\end{itemize}
\noindent where \(\omega^* \in \mathbb{L}(\mathcal{C}^*)\)
is the weighting with \(\omega^*(D) = \omega(D)\)
for all \(D \in \mathcal{C}^*\).

With the properties (NC), (DD) and (DC) in hand, we now present our
characterization of the fair proportion index.

\begin{theorem}
\label{thm:fp:continuous}
Suppose we have, for each cluster system~\(\mathcal{C}\) on~\(X\),
a phylogenetic diversity index \(\Phi_{\mathcal{C}}\) on \(\mathcal{C}\). 
Then the following are equivalent:
\begin{itemize}
\item[(i)]
For all cluster systems \(\mathcal{C}\) on \(X\),
\(\Phi_{\mathcal{C}}\) is the fair proportion index on \(\mathcal{C}\).
\item[(ii)]
For all cluster systems \(\mathcal{C}\) on \(X\),
\(\Phi_{\mathcal{C}}\) is a descendant diversity index that
satisfies the neutrality condition and is downward continuous
with respect to \(\Phi_{\mathcal{C} \setminus \{C\}}\) for all
\(C \in \mathcal{C}\) such that
\(\mathcal{C} \setminus \{C\} \neq \emptyset\).
\end{itemize}
\end{theorem}

\pf
We first show that~(i) implies~(ii).
Consider a cluster system \(\mathcal{C}\) on \(X\)
and put \(\Phi = \Phi_{\mathcal{C}}\). By assumption,
\(\Phi\) is the fair proportion index on \(\mathcal{C}\).
Thus, in view of Lemma~\ref{lem:FPproperties},
\(\Phi\) is linear and complete.
Moreover, as illustrated by
the example in Figure~\ref{fig:gamma:fp:ex}(b),
it follows immediately from the definition of the fair
proportion index in~(\ref{def:fp})
that \(\Phi\) is a descendant diversity index
and satisfies the neutrality condition.

It remains to establish downward continuity.
Consider a cluster \(C \in \mathcal{C}\) and assume
that \(\mathcal{C}^* = \mathcal{C} \setminus \{C\} \neq \emptyset\).
Put \(\Phi^* = \Phi_{\mathcal{C}^*}\). Let
\(\Gamma = \Gamma_{\Phi}\) and
\(\Gamma^* = \Gamma_{\Phi^*}\) be the matrices whose entries satisfy
Equation~(\ref{eq:matrix:linear:pdi}) for
\(\Phi\) and \(\Phi^*\), respectively.
By assumption, \(\Phi\)~is the fair proportion
index on \(\mathcal{C}\) and
\(\Phi^*\)~is the fair proportion index on \(\mathcal{C}^*\)
Therefore, it follows again from the definition of the fair
proportion index in~(\ref{def:fp})
that deleting the row corresponding to the cluster~\(C\)
from the matrix~\(\Gamma\) yields the matrix~\(\Gamma^*\).
But this immediately implies that Equation~(\ref{eq:limit:continuous})
holds for all \(\omega \in \mathbb{L}(\mathcal{C})\),
as required.

Next we show that~(ii) implies~(i).
Let~\(\mathcal{C}\) be a cluster system on~\(X\).
By assumption, \(\Phi = \Phi_{\mathcal{C}}\) is a
descendant diversity index and, therefore, linear.
Let \(\Gamma = \Gamma_{\Phi}\) be the matrix
whose entries satisfy Equation~(\ref{eq:matrix:linear:pdi})
for \(\Phi\). 
In view of the definition of the fair
proportion index in~(\ref{def:fp}), it suffices to show that
the entries of~\(\Gamma\) satisfy
\[
\gamma_{(C,x)} =
\begin{cases}
\frac{1}{|C|} &\text{for} \ x \in C\\
0 &\text{for} \ x \not \in C
\end{cases}
\]
for all \(C \in \mathcal{C}\) and all \(x \in X\).
We use induction on \(|\mathcal{C}|\) to show this.

To establish the base case of the induction, assume
\(|\mathcal{C}| = 1\). Consider \(C \in \mathcal{C}\) and \(x \in X\).
In view of \(|\mathcal{C}| = 1\) we have \(ch(C) = \emptyset\).
Thus, in view of the assumption that
\(\Phi\) is a descendant diversity index and satisfies the
neutrality condition, we have
\(\gamma_{(C,x)}=\frac{1}{|C|}\) for all \(x \in C\) and
\(\gamma_{(C,x)}=0\) for all \(x \in X \setminus C\),
as required.

Next assume \(|\mathcal{C}| \geq 2\).
Consider \(C \in \mathcal{C}\) and put
\(\mathcal{C}^* = \mathcal{C} \setminus \{\mathcal{C}\}\).
By the assumption that \(\Phi\) is downward
continuous with respect to \(\Phi^* = \Phi_{\mathcal{C}^*}\), the matrix
\(\Gamma^* = \Gamma_{\Phi^*}\) whose entries satisfy
Equation~(\ref{eq:matrix:linear:pdi}) for
\(\Phi^*\) is obtained by deleting the
row corresponding to cluster~\(C\) from~\(\Gamma\).
Thus, by induction, we have
\[
\gamma_{(D,x)} =
\begin{cases}
\frac{1}{|D|} &\text{for} \ x \in D\\
0 &\text{for} \ x \not \in D
\end{cases}
\]
for all \(D \in \mathcal{C} \setminus \{C\}\) and all \(x \in X\).
Since this holds for all \(C \in \mathcal{C}\), this
finishes the inductive proof.
\epf

\section{The Shapely value}
\label{sec:shapely}

The Shapely value is a well-known phylogenetic diversity index
that can be computed using rooted phylogenetic trees and
that has its origins in game theory. Interestingly, to understand
a generalization of this index in the cluster setting, it 
is necessary to consider mappings on slightly more general spaces
than those used in the definition of cluster-based phylogenetic
diversity indices in Section~\ref{sec:clusters}.
In this section, we shall explain this, and then
give a characterization of a cluster-based version of the Shapely value.

Let \(X\) be a finite set.
A \emph{game} is a map
\(g : \mathcal{P}(X) \rightarrow \mathbb{R}\).
The elements of \(X\) are referred to as the \emph{players}
in this context and the value \(g(C)\) for some
\(C \in \mathcal{P}(X)\) can be interpreted as
the gain when the players in \(C\) form a coalition.
The aim of the game is then to quantify, for each player \(x \in X\),
the value \(v(x) \in \mathbb{R}\) of the player
with respect to a given game (see, e.g., \citealt{branzei2008models}
for a more detailed exposition of these concepts). 
Formally speaking, we are thus interested in maps \(v\) from
\(\mathbb{R}^{\mathcal{P}(X)}\) to \(\mathbb{R}^X\), and
the \emph{Shapley value} is one specific such map~\(v\) given by 
\begin{equation}
\label{equation:SV}
(v(g))(x) = \frac{1}{|X|!} \cdot  \sum_{M \in \mathcal{P}(X): \ x \in M} 
	[(|M|-1)! \cdot (|X|-|M|)! \cdot (g(M)-g(M \setminus \{x\}))].
\end{equation}
This map was originally proposed by \citet{sha-53a}.
 
In a biological context, the players of Shapley's game are species and
from a rooted phylogenetic tree~\(\mathcal{T}\) on~\(X\)
with edge weights we obtain a game $g$ by setting $g(M)=PD(M)$ for each
\(M \in \mathcal{P}(X)\),  where~$PD(M)$ is the \emph{phylogenetic diversity}
of~\(M\). The value~$PD(M)$ is defined as the total weight
of those edges in~\(T\) that lie on a path from the root
to some species in~\(M\) \citep{fai-92}. For example, for the
rooted phylogenetic tree in Figure~\ref{fig:fp:ex:rooted:unrooted}(a)
we obtain
\[PD(\{a,b,d\}) = 3+2+4+3+2+1 = 15.\]

We now explain a way to generalize these considerations to 
cluster systems~\(\mathcal{C}\) on~\(X\). First we need to 
define the phylogenetic diversity of a subset of~$X$
relative to a weighted cluster system.
Let \(\omega \in \mathbb{L}(\mathcal{C})\). Then 
the \emph{phylogenetic diversity} of a subset \(M\) of \(X\)
with respect to \(\omega\) is defined as 
\begin{equation}
PD(M) = PD_{\omega}(M) = \sum_{C \in \mathcal{C}: \ M \cap C \neq \emptyset} \omega(C).
\label{eq:rooted:pd}
\end{equation}
To further explore properties of the
Shapley value in the context of our work,
it will be convenient to consider the set 
\[\mathbb{PD}(\mathcal{C}) = \{g \in \mathbb{R}^{\mathcal{P}(X)} :
\ \text{there exists} \ \omega \in \mathbb{L}(\mathcal{C}) \
\text{with} \ g = PD_{\omega}\},\]
that is, the set of games in  $\mathbb{R}^{\mathcal{P}(X)}$ 
for which there is some $\omega \in \mathbb{L}(\mathcal{C})$
which gives rise to this game.
The next technical lemma summarizes the
key structural properties of this set.

\begin{lemma}
\label{lem:dim:pd:c}
Let \(\mathcal{C}\) be a cluster system on~\(X\). Then
\(\mathbb{PD}(\mathcal{C})\) is a linear subspace of 
\(\mathbb{R}^{\mathcal{P}(X)}\) that has dimension
\(|\mathcal{C}|\).
\end{lemma}

\pf 
For \(C \in \mathcal{P}(X)\), 
let \(g_C: \mathcal{P}(X) \rightarrow \mathbb{R}\) be the game defined
by putting
\begin{equation}
\label{eq:weighting:omega:c}
g_C(M) = 
\begin{cases}
1 &\text{if} \ C \cap M \neq \emptyset\\
0 &\text{if} \ C \cap M = \emptyset.
\end{cases}
\end{equation}
In view of~(\ref{eq:rooted:pd}) and~(\ref{eq:weighting:omega:c}),
\(\mathbb{PD}(\mathcal{C})\) is the linear span of
the games \(g_C\) for \(C \in \mathcal{C}\):
\[PD_{\omega}(M) = \sum_{C \in \mathcal{C}: \ M \cap C \neq \emptyset} \omega(C)
= \sum_{C \in \mathcal{C}} \omega(C) \cdot g_{C}(M)\]
Thus, it suffices to show that the games \(g_C\),
\(C \in \mathcal{C}\), are linearly independent. 
To see this, consider the square matrix \(A\) whose rows and
columns are each in one-to-one correspondence
with the elements of \(\mathcal{P}(X) \setminus \{\emptyset\}\).
For all \(C,M \in \mathcal{P}(X) \setminus \{\emptyset\}\)
the entry of \(A\) in the row corresponding to \(C\) and
the column corresponding to \(M\) is~1
if \(C \cap M \neq \emptyset\) and is~0 otherwise.
\(A\)~is the so-called \emph{intersection matrix}
of \(\mathcal{P}(X) \setminus \{\emptyset\}\) and it is known
that \(A\) has full rank (see, e.g., \citealt[p. 216]{juk-11a}).
Thus, in particular, the rows corresponding to \(C \in \mathcal{C}\)
are linearly independent.
\epf

Now, as explained above, for a cluster system~\(\mathcal{C}\) on~\(X\),
we restrict in~(\ref{equation:SV})
to games~\(g=PD\) in~\(\mathbb{PD}(\mathcal{C})\).
More specifically, we define the Shapley value
\emph{relative to the cluster system} $\mathcal{C}$
as the map \(SV: \mathbb{PD}(\mathcal{C}) \rightarrow \mathbb{R}^X\)
obtained by putting
\begin{equation}
\label{equation:formula:shapley:value:intro}
(SV(PD))(x) = \frac{1}{|X|!} \cdot \sum_{M \in \mathcal{P}(X): \ x \in M} \left [ (|M|-1)! \cdot (|X|-|M|)! \cdot (PD(M)-PD(M \setminus \{x\})) \right ]
\end{equation}
for all \(PD \in \mathbb{PD}(\mathcal{C})\) and all \(x \in X\).
Note that, in view of Lemma~\ref{lem:dim:pd:c},
\(\mathbb{PD}(\mathcal{C})\) may be a proper subspace
of~\(\mathbb{R}^{\mathcal{P}(X)}\) (i.e. the set of all games).
As we will see below,
any characterization of the Shapley value relative 
to a cluster system must take this into account (see also~\citealt{dub-75a}
for a more general discussion of this aspect). 

The sharp-eyed reader will have noticed that 
the Shapley value relative to a cluster system~\(\mathcal{C}\)
is \emph{not} a phylogenetic diversity index on~\(\mathcal{C}\),
as the latter is defined as a map from
\(\mathbb{L}(\mathcal{C})\) to \(\mathbb{R}^X\).
However, we can resolve this issue by slightly generalizing 
our cluster-based definition of phylogenetic diversity indices.
Let~\(\mathbb{L}\) be a linear subspace of
\(\mathbb{R}^{\mathcal{P}(X)}\). Then we define a
phylogenetic diversity index \emph{on}~\(\mathbb{L}\)
to be a map \(\Phi : \mathbb{L} \rightarrow \mathbb{R}^X\).
This encompasses then the Shapley value as a
phylogenetic diversity index on
\(\mathbb{L} = \mathbb{PD}(\mathcal{C})\) for
all cluster systems \(\mathcal{C}\) on \(X\).
Moreover, viewing \(\mathbb{L}(\mathcal{C})\) as
the linear subspace
\[\mathbb{L} = \{\omega \in \mathbb{R}^{\mathcal{P}(X)} : \omega(C)=0 \
\text{for all} \ C \not \in \mathcal{C}\},\]
it also encompasses phylogenetic diversity indices
on \(\mathcal{C}\) as defined in Section~\ref{sec:clusters}.
In fact, we can say even more about these relationships,
which we will return to in the next section.
 
For the remainder of this section, we focus on
giving a characterization of the Shapley value relative
to a cluster system. This will involve the following two
properties. We say that a  
phylogenetic diversity index~\(\Phi\) on a linear
subspace~\(\mathbb{L}\) of~\(\mathbb{R}^{\mathcal{P}(X)}\) 
satisfies \emph{Pareto efficiency} if
\begin{itemize}
\item[(PE)]
\(\sum_{x \in X} (\Phi(\omega))(x) = \omega(X)\)
for all \(\omega \in \mathbb{L}\).
\end{itemize}

\begin{figure}
\centering
\includegraphics[scale=1.0]{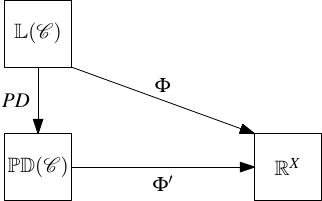}
\caption{This diagram depicts the relationship between a
phylogenetic diversity index
\(\Phi\) on \(\mathbb{L}(\mathcal{C})\) for a cluster system
\(\mathcal{C}\) on \(X\) and a phylogenetic diversity index
\(\Phi'\) on \(\mathbb{PD}(\mathcal{C})\) as
described by Equation~(\ref{eq:commuting:phi:pd:phiprime}).}
\label{fig:diagram:first:part}
\end{figure}

\begin{remark}
\label{rem:link:complete:pareto}
The properties of completeness and Pareto efficiency are
tightly linked. Let \(\mathcal{C}\) be a cluster system on~\(X\)
and note that \(\sum_{C \in \mathcal{C}} \omega(C) = PD_{\omega}(X)\)
holds for all \(\omega \in \mathbb{L}(\mathcal{C})\).
Therefore, every complete phylogenetic diversity index~\(\Phi\)
on~ \(\mathbb{L}(\mathcal{C})\) corresponds to a
phylogenetic diversity index~\(\Phi'\) on~\(\mathbb{PD}(\mathcal{C})\)
that satisfies Pareto efficiency, where \(\Phi'\) is obtained such that
the diagram in Figure~\ref{fig:diagram:first:part} commutes, that is, 
\begin{equation}
\label{eq:commuting:phi:pd:phiprime}
\Phi'(PD_{\omega}) = \Phi(\omega)
\end{equation}
for all \(\omega \in \mathbb{L}(\mathcal{C})\).
\end{remark}

We say that a phylogenetic diversity index \(\Phi\) on a linear
subspace \(\mathbb{L}\) of
\(\mathbb{R}^{\mathcal{P}(X)}\) 
satisfies \emph{group proportionality} (cf. \citealt{haa-kas-08a}) if
\begin{itemize}
\item[(GP)]
\((\Phi(a \cdot g_C))(x) = 
\begin{cases}
\frac{a}{|C|} &\text{if} \ x \in C\\
0 &\text{if} \ x \not \in C,
\end{cases}\)
\quad for all \(C \in \mathcal{P}(X) \setminus \{\emptyset\}\) and all \(a \in \mathbb{R}\)
\end{itemize}
with \(g_C\) the game as defined in~(\ref{eq:weighting:omega:c}).
Note that a similar characterization to that given
in the following theorem
was established by~\citet[Thm.~7]{wic-ste-19a} 
for the special case of cluster systems that form a hierarchy.

\begin{theorem}
\label{thm:characterization:shapley}
Let~\(\mathcal{C}\) be a cluster system on~\(X\).
The Shapley value is the unique phylogenetic diversity index
on \(\mathbb{PD}(\mathcal{C})\)
that is additive and satisfies Pareto efficiency and group
proportionality.
\end{theorem}

\pf
Assume that~\(\Phi'\) is the Shapley value on~\(\mathbb{PD}(\mathcal{C})\).
It is known (see, e.g., \citealt{aum94a})
that~\(\Phi'\) satisfies Pareto efficiency for all
\(\omega \in \mathbb{R}^{\mathcal{P}(X)}\)
and is additive for all
\(\omega_1, \omega_2 \in \mathbb{R}^{\mathcal{P}(X)}\).
Thus, these two properties hold, in particular, for
all \(\omega,\omega_1,\omega_2 \in \mathbb{PD}(\mathcal{C})
\subseteq \mathbb{R}^{\mathcal{P}(X)}\).

To establish that~\(\Phi'\) also satisfies group proportionality,
consider \(x \in X\), \(C \in \mathcal{C}\) and \(a \in \mathbb{R}\).
We calculate the value \((\Phi'(a \cdot g_C))(x)\)
using Formula~(\ref{equation:formula:shapley:value:intro})
(similar calculations are used in the proofs
of \citealt[Thm.~4]{haa-kas-08a} and \citealt[Thm.~1]{cor-rie-18a}):

If \(x \not \in C\) we have
\(g_C(M \cup \{x\})-g_C(M) = 0\) for all \(M \in \mathcal{P}(X)\),
implying \(\Phi'(a \cdot g_C))(x) = 0\), as required.
So assume that \(x \in C\), put \(c = |C|\) and put \(n = |X|\). Then,
in view of the fact that only \(M \in \mathcal{P}(X)\) with
\(M \cap C = \emptyset\) contribute to \((\Phi'(a \cdot g_C))(x)\), we have
\begin{align*}
(\Phi'(a \cdot g_C))(x) &= \frac{a}{n!} \cdot \sum_{m=0}^{n-c} m! \cdot (n-m-1)! \cdot {n-c \choose m} = \frac{a \cdot (n-c)! \cdot (c-1)!}{n!} \cdot \sum_{j=c-1}^{n-1} {j \choose c-1}\\
&= a \cdot \frac{(n-c)! \cdot (c-1)!}{n!} \cdot {n \choose c} = \frac{a}{c},
\end{align*}
as required, where we used the formula for the sum along a diagonal in Pascal's triangle to obtain the first equality in the second line.

Uniqueness now follows from the fact that,
in view of the proof of Lemma~\ref{lem:dim:pd:c},
\(\mathbb{PD}(\mathcal{C})\) is the linear span
of \(\{g_C : C \in \mathcal{C}\}\).
\epf

Interestingly, as shown by \citet{fuc-jin-15a},
the vector in \(\mathbb{R}^X\) that results from computing
the Shapley value on the game \(PD\) obtained from
an edge-weighted rooted phylogenetic tree
always coincides with the
vector that we obtain by computing the fair proportion index
on the rooted phylogenetic tree.
In fact, this is a particular
instance of~(\ref{eq:commuting:phi:pd:phiprime}).
The following Corollary of Theorem~\ref{thm:characterization:shapley}
makes this more precise.

\begin{corollary}
\label{corollary:fp:equals:sv}
Let~\(\mathcal{C}\) be a cluster system on~\(X\),
\(\Phi\) be the fair proportion index on \(\mathbb{L}(\mathcal{C})\), and
\(\Phi'\) be the Shapley value on \(\mathbb{PD}(\mathcal{C})\).
Then
\[\Phi(\omega) = \Phi'(\sum_{C \in \mathcal{C}} \omega(C) \cdot g_C) =
\Phi'(PD_{\omega})\]
holds for all \(\omega \in \mathbb{L}(\mathcal{C})\).
\end{corollary}

\pf
This follows immediately from the definition of the fair proportion index
together with the fact that, by Theorem~\ref{thm:characterization:shapley},
the Shapley value is additive and
satisfies group proportionality.
\epf

It is remarked in the discussion by \citet{cor-rie-18a}
that Corollary~\ref{corollary:fp:equals:sv} can also be derived using 
arguments based on so-called phylogenetic networks (for more on
the connection between such networks and diversity indices see 
Section~\ref{sec:conclusion}). Moreover, 
the fact that the Shapley value 
on \(\mathbb{PD}(\mathcal{C})\) satisfies
Pareto efficiency means that it apportions the phylogenetic
diversity of~$X$ among the elements of~$X$.
In view of Corollary~\ref{corollary:fp:equals:sv}
this then also holds for the fair proportion index
on~\(\mathbb{L}(\mathcal{C})\) and, in view of
Remark~\ref{rem:link:complete:pareto}, this corresponds
to the fact that the fair proportion index 
is complete, as can be seen
in the example in Figure~\ref{fig:gamma:fp:ex}(a):
\[PD_{\omega}(X) = \sum_{C \in \mathcal{C}} \omega(C) = 25
= \sum_{x \in X} (FP(\omega))(x).\]

\section{An affine and projective framework for phylogenetic diversity indices}
\label{sec:framework}

As mentioned in the introduction, the notion of
phylogenetic diversity indices has also
been considered on unrooted phylogenetic
trees \citep{haa-kas-08a,wic-ste-19a} and, just as rooted
phylogenetic trees can be encoded by
a collection of clusters, unrooted phylogenetic
trees on a set~\(X\) of species can be encoded
by a collection~\(\mathcal{S}\) of bipartitions, or splits,
of $X$ (see, e.g., \citealt[Ch.~2]{ste-16}).
In the area of phylogenetic combinatorics, the 
interplay between collections of clusters and collections of 
splits has been studied in terms of affine and
projective models of clustering, respectively, in analogy with the interplay
between affine and projective geometry in classical geometry
(\citealt[p.~207]{dress2012basic}; see also \citealt{dre-97a}).
One of the key ideas that we will exploit from this theory
is that we can map a collection~\(\mathcal{S}\) of splits 
of~\(X\) in a natural way to a cluster system~$\mathcal{C}(\mathcal{S})$
on~\(X\) (defined in~(\ref{eq:c:s}) below)
and, in this way, derive split-based indices
from cluster-based indices.
In this section, we will make this more precise,
and illustrate the resulting framework using the fair proportion index and
the Shapely value as examples. 

First, we formally define the concepts mentioned above.
A \emph{split} \(S\) of \(X\) is a bipartition
of \(X\) into two non-empty subsets \(A\) and \(B\), that is,
\(A \cup B = X\) and \(A \cap B = \emptyset\). We denote such
a split as an unordered pair \(A|B = B|A\).
A \emph{split system}~\(\mathcal{S}\) on~\(X\) is a non-empty set
of splits of~\(X\). By \(\mathcal{S}(X)\) we denote the set of
all splits of~\(X\) and, for a split system
\(\mathcal{S} \subseteq \mathcal{S}(X)\), we denote
by \(\mathbb{L}(\mathcal{S})\) the set of all weightings
\(\lambda : \mathcal{S}(X) \rightarrow \mathbb{R}\)
with \(\lambda(S) = 0\) for all \(S \in \mathcal{S}(X) \setminus \mathcal{S}\).
In addition, we denote by \(\mathbb{PD}(\mathcal{S})\) the 
set of all weightings
\(PD : \mathcal{P}(X) \rightarrow \mathbb{R}\)
that can be written as
\begin{equation}
\label{eq:def:pd:splits}
PD(M) = PD_{\lambda}(M) = \sum_{A|B \in \mathcal{S}: \ A \cap M \neq \emptyset, B \cap M \neq \emptyset} \lambda(A|B)
\end{equation}
for some \(\lambda \in \mathbb{L}(\mathcal{S})\).
The value \(PD_{\lambda}(M)\) is usually called the
\emph{phylogenetic diversity} of~\(M\) with respect to
the weighting~\(\lambda\) of the splits in~\(\mathcal{S}\)
(see, e.g., \citealt{spi-ngu-08a}). 

\begin{figure}
\centering
\includegraphics[scale=1.0]{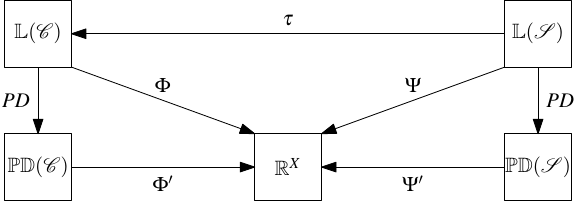}
\caption{A diagram of the various maps we consider to
study relationships between phylogenetic diversity indices.
The left part of the diagram we have already seen in
Figure~\ref{fig:diagram:first:part}.
In analogy to this, the right part of the
diagram depicts phylogenetic diversity
indices \(\Psi\) and \(\Psi'\) on~\(\mathbb{L}(\mathcal{S})\)
and~\(\mathbb{PD}(\mathcal{S})\), respectively, where~\(\mathcal{S}\) is a
split system on~\(X\). Finally~\(\tau\) associates with
each weighting~\(\lambda\) of the splits in~\(\mathcal{S}\)
a weighting~\(\omega = \tau(\lambda)\) of the clusters
in a cluster system~\(\mathcal{C} = \mathcal{C}(\mathcal{S})\)
that arises from~\(\mathcal{S}\) by~(\ref{eq:c:s}).}
\label{fig:diagram:maps}
\end{figure}

Figure~\ref{fig:diagram:maps} gives an overview of the various spaces
we shall consider and the maps between them.
In addition to the maps already introduced in 
in Figure~\ref{fig:diagram:first:part}
in Section~\ref{sec:shapely}, we also consider,
for split systems \(\mathcal{S}\) on \(X\),
maps \(\tau\) from \(\mathbb{L}(\mathcal{S})\) to
\(\mathbb{L}(\mathcal{C})\) where \(\mathcal{C}\) is
the cluster system
\begin{equation}
\label{eq:c:s}
\mathcal{C}(\mathcal{S}) = \bigcup_{S \in \mathcal{S}} S
\end{equation}
on~\(X\) mentioned above.
In particular, we are interested in maps \(\tau\) for 
which various parts of the diagram in
Figure~\ref{fig:diagram:maps} commute. 

As an illustration
of this setup, we now revisit the relationship between the
fair proportion index and the Shapely value. 
Let \(\mathcal{S}\) be a split system on~\(X\).
Then the \emph{Shapley value} on \(\mathbb{PD}(\mathcal{S})\) is
defined as in~(\ref{equation:formula:shapley:value:intro}).
Equivalently, as shown by~\citet{haa-kas-08a} for trees
and by~\citet{vol-mar-14a} for split systems in general, the Shapley
value on \(\mathbb{PD}(\mathcal{S})\) can also be computed as
\begin{equation}
\label{eq:alter:sv}
(SV(PD_{\lambda}))(x) =
\sum_{A|B \in \mathcal{S}: \ x \in A} \frac{|B|}{|X| \cdot |A|} \cdot \lambda(A|B). 
\end{equation}
for all \(\lambda \in \mathbb{L}(\mathcal{S})\) and all \(x \in X\).

\begin{figure}
\centering
\includegraphics[scale=1.0]{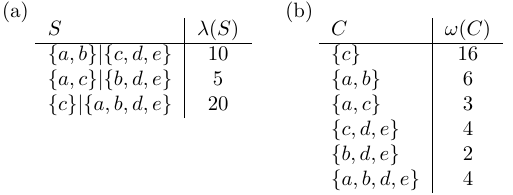}
\caption{(a) A split system \(\mathcal{S}\) on \(X=\{a,b,c,d,e\}\)
with weighting \(\lambda\).
(b)~The associated cluster system \(\mathcal{C}(\mathcal{S})\) on~\(X\)
as defined in~(\ref{eq:c:s}) and the weighting \(\omega = \tau(\lambda)\)
as defined in~(\ref{eq:tau:1}).}
\label{fig:tau:1:example}
\end{figure}

Now consider the map $\tau: \mathbb{L}(\mathcal{S}) \to  \mathbb{L}(\mathcal{C}(\mathcal{S}))$
defined by putting, for $\lambda \in \mathbb{L}(\mathcal{S})$, 
\begin{equation}
\label{eq:tau:1}
(\tau(\lambda))(A) = \frac{|B|}{|X|} \cdot \lambda(A|B)
\ \text{ and } \ (\tau(\lambda))(B) = \frac{|A|}{|X|} \cdot \lambda(A|B)
\end{equation}
for all \(A,B \in \mathcal{C}(\mathcal{S})\) such that \(A|B\) is a split
in~\(\mathcal{S}\).
For example, consider the split system \(\mathcal{S}\)
with weighting~\(\lambda\) in Figure~\ref{fig:tau:1:example}(a).
Using Formula~(\ref{eq:alter:sv}), we obtain \(SV(a) = \frac{11}{2}\)
in this example and we also have
\(FP(a) = \frac{11}{2}\) for the fair
proportion index as defined in~(\ref{def:fp})
applied to the cluster system \(\mathcal{C}(\mathcal{S})\)
with weighting \(\omega = \tau(\lambda)\).
We conclude this section by showing that this is not a coincidence. 

\begin{theorem}
\label{thm:splits:to:clusters}
Let \(\mathcal{S}\) be a split system on~\(X\),
\(\Phi\) be the fair proportion index
on~\(\mathbb{L}(\mathcal{C}(\mathcal{S}))\) and
\(\Psi'\) be the Shapley value on~\(\mathbb{PD}(\mathcal{S})\).
If $\tau$ is as defined in (\ref{eq:tau:1}), then 
\begin{equation}
\label{eq:cluster:fp:corresponds:to:split:sv}
\Phi(\tau(\lambda)) = \Psi'(PD_{\lambda})
\end{equation}
for all $\lambda \in \mathbb{L}(\mathcal{S})$.
\end{theorem}

\pf
Let $\lambda \in \mathbb{L}(\mathcal{S})$ and put \(\omega = \tau(\lambda)\).
Since the maps \(\Phi\), \(\tau\), \(\Psi'\) and \(PD\)
are all linear, it suffices to show
Equation~(\ref{eq:cluster:fp:corresponds:to:split:sv}) for the case
that one element of $\cS$, say $S=A|B$ has weight~1
(i.e. $\lambda(A|B)=1$), whereas $\lambda(S')=0$ for
all $S' \neq S$. Then we have
$\omega(A) = |B|/|X|$, $\omega(B)=|A|/|X|$, and $\omega(C)=0$
for all $C \in \cC(\cS)$ with $C \neq A,B$.
Now let $x \in X$, and assume without loss of generality that
$x \in A$. Then,
$$ (\Phi(\omega))(x) = \sum\limits_{C \in \cC(\cS): \ x \in C} \frac{\omega(C)}{|C|} = \frac{\omega(A)}{|A|} = \frac{\frac{|B|}{|X|}}{|A|} = \frac{|B|}{|X| \cdot |A|}.$$
On the other hand, in view of~(\ref{eq:alter:sv}) we have
$$ (\Psi'(PD_{\lambda}))(x) = \sum\limits_{A'|B' \in \cS: \ x \in A'} \frac{|B'|}{|X| \cdot |A'|} \lambda(S) = \frac{|B|}{|X| \cdot |A|}$$
as well. This completes the proof.
\epf

\section{Complete diversity indices}
\label{sec:equal:splits}

In this section we shall consider Figure~\ref{fig:diagram:maps}
once again, considering an alternative definition for the 
map \(\tau\) that can be used to translate,
for any split system~\(\mathcal{S}\) on~\(X\), the property
of completeness from a cluster-based 
index \(\Phi\) on  \(\mathbb{L}(\mathcal{C}(\mathcal{S}))\) to an
associated split-based index \(\Psi = \Psi_{\tau}(\Phi)\)
on~\(\mathbb{L}(\mathcal{S})\).
In particular, we will see that
this immediately implies the completeness
of the fair proportion index on unrooted phylogenetic trees
that was established by \citet{wic-ste-19a} (for example, see  
Figure~\ref{fig:fp:ex:rooted:unrooted}(b) in the introduction).
In addition, we illustrate the application of
these considerations to a generalization of the 
so-called equal splits index that appears in \citep{wic-ste-19a}.

We begin by proving a result concerning completeness.
Let~\(\mathcal{S}\) be a split system on~\(X\). 
A phylogenetic diversity index~\(\Psi\) on
\(\mathbb{L}(\mathcal{S})\) is \emph{complete} if
\begin{itemize}
\item[(C')]
\(\sum_{x \in X} (\Psi(\lambda))(x) = \sum_{S \in \mathcal{S}} \lambda(S)\)
holds for all \(\lambda \in \mathbb{L}(\mathcal{S})\).
\end{itemize}
Define the map
$\tau:\mathbb{L}(\mathcal{S}) \to \mathbb{L}(\mathcal{C(\mathcal{S})})$
by putting
\begin{equation}
\label{eq:def:tau:2}
(\tau(\lambda))(C) = \frac{1}{2} \cdot \lambda(C|(X-C))
\end{equation}
for all \(C \in \mathcal{C}(\mathcal{S})\). Then,
for a phylogenetic diversity index \(\Phi\) on
\(\mathbb{L}(\mathcal{C}(\mathcal{S}))\), we define the
phylogenetic diversity index \(\Psi = \Psi_{\tau}(\Phi)\)
on \(\mathbb{L}(\mathcal{S})\) by putting
\(\Psi(\lambda) = \Phi(\tau(\lambda))\)
for all \(\lambda \in \mathbb{L}(\mathcal{S})\).  

\begin{theorem}
\label{theorem:transform:cluster:index:to:split:index}
Let \(\mathcal{S}\) be a split system on \(X\) and 
\(\Phi\) a complete linear phylogenetic diversity index
on \(\mathbb{L}(\mathcal{C}(\mathcal{S}))\).
If $\tau$ is as defined in (\ref{eq:def:tau:2}),
then \(\Psi_{\tau}(\Phi)\) is a complete linear
phylogenetic diversity index
on \(\mathbb{L}(\mathcal{S})\).
\end{theorem}

\pf
Let \(\Phi\) be a complete linear phylogenetic diversity index
on \(\mathbb{L}(\mathcal{C}(\mathcal{S}))\).
We first show that \(\Psi_{\tau}\) is linear.
Let \(\lambda_1, \lambda_2 \in \mathbb{L}(\mathcal{S})\)
and \(a \in \mathbb{R}\). Then, noting that \(\tau\) is linear, we have
\begin{align*}
(\Psi_{\tau}(\Phi))(a \cdot \lambda_1 + \lambda_2)
&= \Phi(\tau(a \cdot \lambda_1 + \lambda_2)) = \Phi(a \cdot \tau(\lambda_1) + \tau(\lambda_2)) \\
&= a \cdot \Phi(\tau(\lambda_1)) + \Phi(\tau(\lambda_2)) = a \cdot (\Psi_{\tau}(\Phi))(\lambda_1) + (\Psi_{\tau}(\Phi))(\lambda_2),
\end{align*}
as required.

It remains to show that \(\Psi_{\tau}\) is complete.
Let \(\lambda \in \mathbb{L}(\mathcal{S})\). Then we have
\begin{align*}
\sum_{x \in X} ((\Psi_{\tau}(\Phi))(\lambda))(x) &= \sum_{x \in X} (\Phi(\tau(\lambda)))(x) = \sum_{C \in \mathcal{C}(\mathcal{S})} (\tau(\lambda))(C)\\
&= \sum_{C \in \mathcal{C}(\mathcal{S})} \frac{1}{2} \cdot \lambda(C|X-C) = \sum_{S \in \mathcal{S}} \lambda(S),
\end{align*}
as required.
\epf

The following Corollary~\ref{cor:fp:splits:complete}
includes, as a special case,
the completeness of the fair proportion index on
unrooted phylogenetic trees that was established by
\citet[Thm.~10]{wic-ste-19a}. To see this, it suffices
to consider, for an unrooted phylogenetic tree on \(X\),
the split system consisting of those splits of~\(X\)
that can be obtained by removing an edge from the tree.

\begin{corollary}
\label{cor:fp:splits:complete}
Let \(\mathcal{S}\) be a split system on \(X\) and
\(\Phi\) be the fair proportion index 
on \(\mathbb{L}(\mathcal{C}(\mathcal{S}))\).
If $\tau$ is as defined in (\ref{eq:def:tau:2}), then
\(\Psi_{\tau}(\Phi)\) is a complete linear phylogenetic
diversity index on \(\mathbb{L}(\mathcal{S})\) and we have
\begin{equation}
\label{eq:fp:splits}
((\Psi_{\tau}(\Phi))(\lambda))(x)
= \sum_{A|B \in \mathcal{S}: \ x \in A} \frac{\lambda(S)}{2 \cdot |A|}
\end{equation}
for all \(\lambda \in \mathbb{L}(\mathcal{S})\) and all \(x \in X\).
\end{corollary}

\pf
In view of Lemma~\ref{lem:FPproperties},
Theorem~\ref{theorem:transform:cluster:index:to:split:index}
implies that \(\Psi_{\tau}(\Phi)\) is a complete linear phylogenetic
diversity index on \(\mathbb{L}(\mathcal{S})\).
Moreover, (\ref{eq:fp:splits}) follows from~(\ref{def:fp}),
(\ref{eq:c:s}), and (\ref{eq:def:tau:2}).
\epf

\begin{figure}
\centering
\includegraphics[scale=1.0]{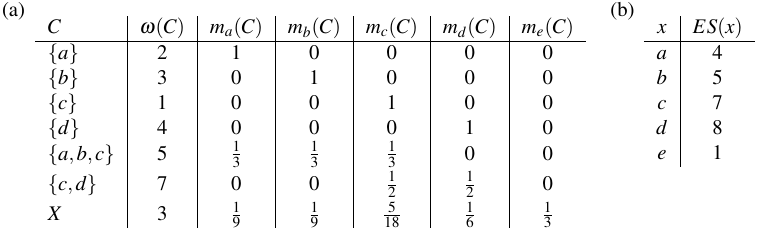}
\caption{(a)~A cluster system \(\mathcal{C}\) on~\(X = \{a,b,c,d,e\}\)
with weighting~\(\omega\) and the quantities \(m_x(C)\)
as defined for all \(x \in X\) and \(C \in \mathcal{C}\)
in~(\ref{eq:mx:def:es}). 
(b)~The equal splits index \(ES(x)\) for all \(x \in X\)
obtained from~\(\mathcal{C}\) and \(\omega\)
by~(\ref{eq:def:es}).}
\label{fig:es:ex}
\end{figure}

We now turn our attention to a generalization of the \emph{equal splits}
index, a phylogenetic diversity index that was introduced in the setting of
rooted phylogenetic trees by \citet{red-moo-06a}.
We first define our generalization for
cluster systems $\cC$ on $X$. 
For all \(C \in \mathcal{C}\),
let $cl(C)$ denote the set of those $x \in C$ that are 
not contained in any cluster $C' \in ch(C)$. Then put
\begin{equation}
\label{eq:mx:def:es}
m_x(C) =  
\begin{cases}
0 &\text{if} \ x \notin C, \\
\frac{1}{|ch(C)|+|cl(C)|} &\text{if} \ x \in cl(C), \\
\sum\limits_{C' \in ch(C)} \frac{m_x(C')}{|ch(C)| + |cl(C)|} &\text{otherwise,}
\end{cases}
\end{equation}
for all \(x \in X\) and all \(C \in \mathcal{C}\).
Note that $m_x(C)=1/|C|$ if $x \in C$ and
$ch(C)=\emptyset$ (as in this case $|ch(C)|=0$ and $|cl(C)|=|C|$).
The equal splits index is defined by putting
\begin{equation}
\label{eq:def:es}
(ES(\omega))(x) = \sum_{C \in \cC} m_x(C) \cdot \omega(C)
\end{equation}
for all $\omega \in \mathbb{L}(\cC)$ and all $x \in X$.
As an example, consider the cluster system \(\mathcal{C}\) 
with weighting~\(\omega\) in
Figure~\ref{fig:es:ex}(a). For the cluster \(C=X\) we have
\(ch(X) = \{\{a,b,c\},\{c,d\}\}\) and \(cl(X) = \{e\}\),
which yields, by~(\ref{eq:mx:def:es}), \(m_e(X) = \frac{1}{3}\).
The resulting values of the equal splits index are
given in Figure~\ref{fig:es:ex}(b).

The equal splits index on \(\mathbb{L}(\mathcal{C})\) is linear with
the corresponding \(|\mathcal{C}| \times |X|\)-matrix~\(\Gamma\)
in Equation~(\ref{eq:matrix:linear:pdi})
having the entries \(\gamma_{(C,x)} = m_x(C)\).
Moreover, as can be seen in the example in Figure~\ref{fig:es:ex}(a), the
sum of the entries in each row of~\(\Gamma\) equals~1. 
The next theorem establishes that this is always the case.

\begin{theorem}
\label{thm:ES:complete}
For all cluster systems~\(\mathcal{C}\) on~\(X\)
the equal splits index is a complete linear phylogenetic diversity index
on~\(\mathbb{L}(\mathcal{C})\).
\end{theorem}

\pf
Let $\cC$ be a cluster system on~\(X\). We already noted above that
the equal splits index is linear. 
Thus, it remains to establish that the equal splits index is complete.
More specifically, it suffices to show that $\sum_{x \in X} m_x(C)=1$
for all $C \in \cC$. We show this by induction on $|ch(C)|$.
For the base case $|ch(C)|=0$ we have
$|cl(C)|=|C|$ and
\[\sum\limits_{x \in X} m_x(C) =  \sum\limits_{x \in C} m_x(C) +  \sum\limits_{x \in X \setminus C} m_x(C) 
= |C| \cdot \frac{1}{|C|} + 0 = 1,\]
as required.
Next assume $|ch(C)|>0$. Then we have
\begin{align*}
\sum\limits_{x \in X} m_x(C) &= \sum\limits_{x \in X \setminus C} m_x(C) + \sum\limits_{x \in cl(C)} m_x(C) + \sum\limits_{x \in C \setminus cl(C)} m_x(C) \\
&= 0 + \frac{|cl(C)|}{|ch(C)|+|cl(C)|} + \sum\limits_{x \in C \setminus cl(C)} \ \sum\limits_{C' \in ch(C)} \frac{m_x(C')}{|ch(C)| + |cl(C)|}\\
&= \frac{|cl(C)|}{|ch(C)|+|cl(C)|} + \sum\limits_{C' \in ch(C)} \ \sum\limits_{x \in C \setminus cl(C)} \frac{m_x(C')}{|ch(C)| + |cl(C)|}\\
&= \frac{|cl(C)|}{|ch(C)|+|cl(C)|} + \sum\limits_{C' \in ch(C)} \ \sum\limits_{x \in X} \frac{m_x(C')}{|ch(C)| + |cl(C)|} \\
&= \frac{|cl(C)|}{|ch(C)|+|cl(C)|} + |ch(C)| \cdot  \frac{1}{|ch(C)| + |cl(C)|}\\
&= 1,
\end{align*}
where the equality in the fourth line holds in view
of the fact that \(m_x(C') = 0\) for all
\(x \in X \setminus (C \setminus cl(C))\) and
for all \(C' \in ch(C)\),
and the equality in the fifth line holds by induction.
\epf

Our final result in this section, which is an immediate consequence
of Theorem~\ref{theorem:transform:cluster:index:to:split:index}
and Theorem~\ref{thm:ES:complete}, summarizes how
we obtain, via the map~$\tau$ defined in~(\ref{eq:def:tau:2}),
a complete linear split-based phylogenetic diversity
index from the cluster-based equal splits index.

\begin{corollary}
\label{cor:fp:and:es:for:split:systems}
Let \(\mathcal{S}\) be a split system on~\(X\) and \(\Phi\)
be the equal splits index on \(\mathbb{L}(\mathcal{C}(\mathcal{S}))\).
If $\tau$ is as defined in (\ref{eq:def:tau:2}), then \(\Psi_{\tau}(\Phi)\) is a complete linear phylogenetic diversity
index on \(\mathbb{L}(\mathcal{S})\).
\end{corollary}

As a concrete example of the last corollary, we
present a computation of the  phylogenetic diversity
index~\(\Psi_{\tau}(\Phi)\).  We 
consider a data set comprising 32~populations of
spotted owls (\emph{Strix occidentalis}) that was
analyzed by~\citet{vol-mar-14a}. In Table~S1 of~\citep{vol-mar-14a}
pairwise genetic distances between these populations are given.
Here, for illustration purposes, we select a subset
\(X = \{a,f,h,m,r,s\}\) of six of these
populations.

\begin{figure}
\centering
\includegraphics[scale=0.32]{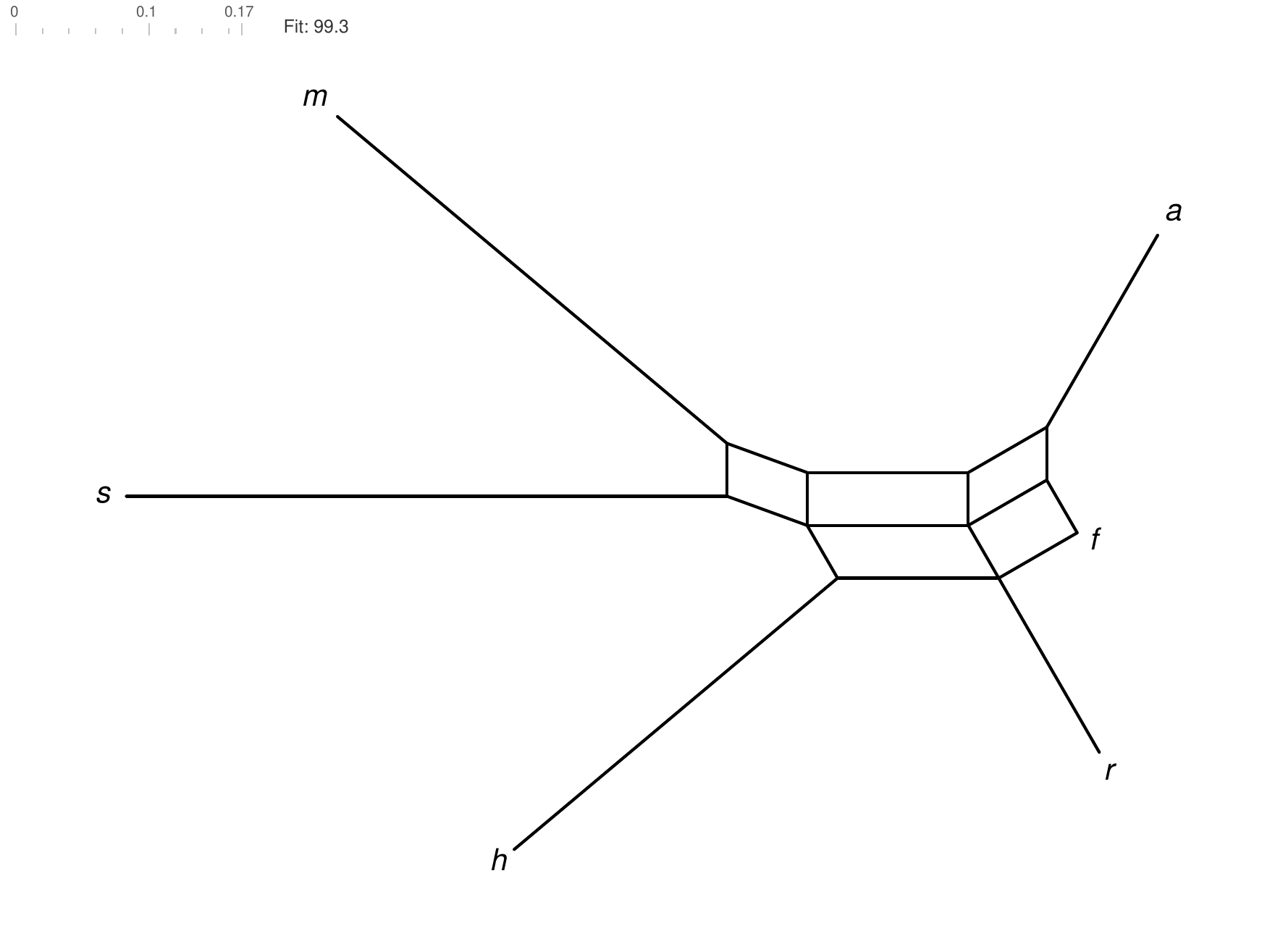}
\caption{A phylogenetic network visualizing a weighted split system
\(\mathcal{S}\) on the set \(X = \{a,f,h,m,r,s\}\) of owl populations
described in the text. The elements
in~\(X\) represent sampling locations of spotted owls
in western North America
($a$ = Aguascalientes, Mexico; $f$ = San Fransisco Peaks, AZ;
$h$ = Huachuca Mountains, AZ; $m$ = Marin County, CA;
$r$ = Capitol Reef National Park, UT; $s$ = San Bernardino Mountains, CA).}
\label{fig:nnet:owls}
\end{figure}

We first compute an unrooted phylogenetic network
from the pairwise distances between them using
the implementation of NeighborNet~\citep{bryant2004neighbor}
in SplitsTree~\citep{huson2005splitstree}
(the same methodology was applied by~\citealt{vol-mar-14a}
to all 32~populations). 
The resulting phylogenetic network is shown
in Figure~\ref{fig:nnet:owls}.
Note that each band of parallel
edges in this network corresponds to a split of the six populations, 
so that the network represents a weighted
split system \(\mathcal{S}\) on~\(X\) that consists
of 10~splits of~\(X\). In addition,  
the length of the edges in the band  
corresponding to a split $S$ gives the weight $\lambda(S)$ of the split.
For example, the band of three 
parallel horizontal edges in the center of the network
in  Figure~\ref{fig:nnet:owls}
corresponds to the split \(S=\{a,f,r\}|\{h,m,s\}\) which
has weight \(\lambda(S) = 0.133\).
The total weight of all splits in~\(\mathcal{S}\) is~\(1.980\)
(all weights rounded to three decimal places).

\begin{figure}
\centering
\includegraphics[scale=1.0]{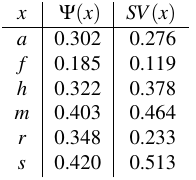}
\caption{The values of the 
phylogenetic diversity indices \(\Psi=\Psi_{\tau}(\Phi)\) and $SV$ for
six owl populations as described in the main text.}
\label{fig:sv:es:owls}
\end{figure}

From the split system~\(\mathcal{S}\) on~\(X\) with
weighting~\(\lambda\) we  compute the cluster system
\(\mathcal{C} = \mathcal{C}(\mathcal{S})\) on~\(X\) with weighting
\(\omega = \tau(\lambda)\) (in Figure~\ref{fig:hasse:owls} in
the Appendix we present
the Hasse diagram for the 20~clusters in~\(\mathcal{C}\), where 
the weight \(\omega(C)\) is given below each cluster~\(C\) in the diagram).
Finally, we compute the  matrix~\(\Gamma = \Gamma_{\Phi}\)
for the equal splits index~\(\Phi\) on~\(\mathbb{L}(\mathcal{C})\)
(see Figure~\ref{fig:gamma:es:owls} in the Appendix) from which
we obtain the values of the phylogenetic diversity
index $\Psi = \Psi_{\tau}(\Phi)$
given in Figure~\ref{fig:sv:es:owls}.
For comparison purposes, we also compute the Shapley value index, $SV$,
based on the NeighborNet\footnote{In fact this 
is the Shapely value of the underlying split system
as derived by \citet{vol-mar-14a}.}  
in Figure~\ref{fig:nnet:owls} using SplitsTreeCE~\citep{huson2005splitstree}.

As can be checked in Figure~\ref{fig:sv:es:owls},
the sum of the values of the index~\(\Psi\)
yields the total weight~\(1.980\) of all splits
in~\(\mathcal{S}\), as it should be for a complete
phylogenetic diversity index. The ranking of the six populations
given by \(\Phi\) and \(SV\) is similar (but not identical):
populations \(s\) and \(m\) are ranked at the top
and population \(f\) is ranked at the bottom by 
both indices. In future work, it could be interesting
to further investigate the differences in rankings obtained from
these and other split-based diversity indices.

\section{Conclusion}
\label{sec:conclusion}

In this paper, we have presented a framework for 
phylogenetic diversity indices defined on linear spaces coming from
weighted cluster and split systems. Using some examples of popular
tree-based phylogenetic diversity indices from the literature,
we have shown that this framework can 
yield generalizations of these indices for cluster and split systems as
well as allowing us to gain a better understanding of their interrelationships.

Note that in our framework presented in 
Figure~\ref{fig:diagram:maps}, 
by associating to any split system \(\mathcal{S}\) on~\(X\)
the cluster system \(\mathcal{C}(\mathcal{S})\) on \(X\) and
then considering maps~\(\tau\),
we have focused on deriving split-based indices from cluster-based indices.
In the affine and projective clustering approach, however, there are 
also ways to associate to any cluster system~\(\mathcal{C}\)
on~\(X\) a split system \(\mathcal{S}(\mathcal{C})\) on~\(X\)
(see, e.g., \citealt[Sec.~9.3]{dress2012basic}).  
Thus, it could be interesting to investigate
how this fact might be used to derive
cluster-based indices from split-based indices,

In our results, we have considered cluster and split systems in general, 
special examples of which include hierarchical cluster systems and compatible 
split systems which correspond
to phylogenetic trees. There are, however, various other 
special classes of 
cluster and split systems
that could be interesting to investigate within our framework.
For example, in \citep{vol-mar-14a} two
specific phylogenetic diversity indices for
so-called \emph{circular split systems} were computed and,
in the light of the 
recent results by \citet{abh-col-23a},
it would be interesting to look at phylogenetic diversity indices
for such split systems more generally. This 
possibility is illustrated 
in the example of six owl populations presented at the end of 
Section~\ref{sec:equal:splits}
since the split systems underlying NeighborNets are always
circular~\citep{bryant2004neighbor}. It would
also be interesting to consider diversity
indices coming from \emph{weak hierarchies},
a special type of cluster system introduced by~\citet{bandelt1989weak}.
The advantage of considering such specialized cluster and split systems
is that they can be efficiently computed from biological data,
making them potentially more useful for applications. 

In the literature, various approaches have been proposed
to generalize tree-based phylogenetic diversity indices
using phylogenetic networks, a graph-theoretical generalization of 
phylogenetic trees \citep{cor-rie-18a,vol-mar-14a,wic-fis-18a}.
Such networks are essentially directed, acyclic, graphs with
a single root and whose set of leaves
corresponds to some collection of species.
The fair proportion index, for example, is generalized 
in terms of such networks by \citet{cor-rie-18a}. In general,
phylogenetic networks give rise to cluster
systems (see, e.g., \citealt[Sec.~10.3.4]{ste-16})
by, for example, taking the set of leaves that lie below 
a vertex or an edge in the network (just as with rooted phylogenetic trees).
Thus, it could be interesting
to explore how phylogenetic diversity indices
defined in terms of phylogenetic networks, such as,
for example, those considered by \citet{wic-fis-18a},
fit into our cluster based framework. This could also
be interesting to investigate for \emph{split networks} such as the 
one presented in Figure~\ref{fig:nnet:owls}, which are 
a certain type of undirected  
phylogenetic network (see, e.g., \citealt[Sec.~4.4]{dress2012basic}).

Finally, in another direction, it could be interesting 
to apply our framework to establish properties and generalizations of other tree-based phylogenetic
diversity indices that we did not consider in this paper.
Indeed, as we have demonstrated, sometimes expressing
indices in terms of clusters or splits can lead to more concise proofs for 
showing that they have certain properties. 
For example, it would be interesting to
extend the Pauplin index
considered by \citet{wic-ste-19a} for unrooted phylogenetic trees
to more general split systems and study its properties, as well 
as to consider some of the other related questions asked
in \citep[Sec.~6]{wic-ste-19a} within our new framework.

\subsection*{Data availability}
No data was generated.

\subsection*{Acknowledgment}
All authors thank Schloss Dagstuhl -- Leibniz Centre for Informatics -- for hosting the Seminar 19443 {\it Algorithms and Complexity in Phylogenetics} in October 2019, where this work was initiated.

\bibliographystyle{apalike}
\bibliography{pd_indices}

\begin{thebibliography}{}

\bibitem[Abhari et~al., 2023]{abh-col-23a}
Abhari, N., Colijn, C., Mooers, A., and Tupper, P. (2023).
\newblock Capturing diversity: split systems and circular approximations for
  conservation.
\newblock {\em Journal of Theoretical Biology}.

\bibitem[Aumann, 1994]{aum94a}
Aumann, R.~J. (1994).
\newblock The {S}hapley value.
\newblock In {\em Game-Theoretic Methods in General Equilibrium Analysis},
  pages 61--66. Springer.

\bibitem[Bandelt and Dress, 1989]{bandelt1989weak}
Bandelt, H.-J. and Dress, A.~W. (1989).
\newblock Weak hierarchies associated with similarity measures—an additive
  clustering technique.
\newblock {\em Bulletin of Mathematical Biology}, 51(1):133--166.

\bibitem[Bordewich and Semple, 2023]{bor-sem-23a}
Bordewich, M. and Semple, C. (2023).
\newblock Quantifying the difference between phylogenetic diversity and
  diversity indices.
\newblock arXiv:2304.10725.

\bibitem[Branzei et~al., 2008]{branzei2008models}
Branzei, R., Dimitrov, D., and Tijs, S. (2008).
\newblock {\em Models in cooperative game theory}, volume 556.
\newblock Springer Science \& Business Media.

\bibitem[Bryant and Moulton, 2004]{bryant2004neighbor}
Bryant, D. and Moulton, V. (2004).
\newblock Neighbor-net: an agglomerative method for the construction of
  phylogenetic networks.
\newblock {\em Molecular Biology and Evolution}, 21(2):255--265.

\bibitem[Coronado et~al., 2018]{cor-rie-18a}
Coronado, T.~M., Riera, G., and Rossell{\'o}, F. (2018).
\newblock The fair proportion is a {S}hapley value on phylogenetic networks
  too.
\newblock In {\em Enjoying Natural Computing}, pages 77--87. Springer.

\bibitem[Dress, 1997]{dre-97a}
Dress, A. (1997).
\newblock Towards a theory of holistic clustering.
\newblock In {\em Mathematical Hierarchies and Biology}, pages 271--289.
  American Mathematical Society.

\bibitem[Dress, 2012]{dress2012basic}
Dress, A. (2012).
\newblock {\em Basic phylogenetic combinatorics}.
\newblock Cambridge University Press.

\bibitem[Dubey, 1975]{dub-75a}
Dubey, P. (1975).
\newblock On the uniqueness of the {S}hapley value.
\newblock {\em International Journal of Game Theory}, 4:131--139.

\bibitem[Faith, 1992]{fai-92}
Faith, D.~P. (1992).
\newblock Conservation evaluation and phylogenetic diversity.
\newblock {\em Biological Conservation}, 61(1):1--10.

\bibitem[Fuchs and Jin, 2015]{fuc-jin-15a}
Fuchs, M. and Jin, E.~Y. (2015).
\newblock Equality of {S}hapley value and fair proportion index in phylogenetic
  trees.
\newblock {\em Journal of Mathematical Biology}, 71(5):1133--1147.

\bibitem[Gumbs et~al., 2023]{gum-gra-23}
Gumbs, R., Gray, C.~L., B\"{o}hm, M., Burfield, I.~J., Couchman, O.~R., Faith,
  D.~P., Forest, F., Hoffmann, M., Isaac, N. J.~B., Jetz, W., Mace, G.~M.,
  Mooers, A.~O., Safi, K., Scott, O., Steel, M., Tucker, C.~M., Pearse, W.~D.,
  Owen, N.~R., and Rosindell, J. (2023).
\newblock The {EDGE}2 protocol: {A}dvancing the prioritisation of
  evolutionarily distinct and globally endangered species for practical
  conservation action.
\newblock {\em {PLOS} Biology}, 21(2):e3001991.

\bibitem[Haake et~al., 2008]{haa-kas-08a}
Haake, C.-J., Kashiwada, A., and Su, F.~E. (2008).
\newblock The {S}hapley value of phylogenetic trees.
\newblock {\em Journal of Mathematical Biology}, 56(4):479--497.

\bibitem[Huson and Bryant, 2005]{huson2005splitstree}
Huson, D.~H. and Bryant, D. (2005).
\newblock Application of phylogenetic networks in evolutionary studies.
\newblock {\em Molecular Biology and Evolution}, 23(2):254–267.

\bibitem[Isaac et~al., 2007]{isa-tur-07}
Isaac, N.~J., Turvey, S.~T., Collen, B., Waterman, C., and Baillie, J.~E.
  (2007).
\newblock Mammals on the {EDGE}: {C}onservation priorities based on threat and
  phylogeny.
\newblock {\em PLoS ONE}, 2(3):e296.

\bibitem[Jukna, 2011]{juk-11a}
Jukna, S. (2011).
\newblock {\em Extremal combinatorics: with applications in computer science}.
\newblock Springer.

\bibitem[Kleinman et~al., 2013]{kleinman2013affine}
Kleinman, A., Harel, M., and Pachter, L. (2013).
\newblock Affine and projective tree metric theorems.
\newblock {\em Annals of Combinatorics}, 17:205--228.

\bibitem[Manson, 2022]{man-22a}
Manson, K. (2022).
\newblock Many phylogenetic diversity indices are not robust to extinctions.
\newblock bioRxiv:2022.06.28.498028.

\bibitem[Manson and Steel, 2023]{man-ste-23a}
Manson, K. and Steel, M. (2023).
\newblock Spaces of phylogenetic diversity indices: combinatorial and geometric
  properties.
\newblock {\em Bulletin of Mathematical Biology}, 85(78).

\bibitem[Redding, 2003]{red-03}
Redding, D.~W. (2003).
\newblock Incorporating genetic distinctness and reserve occupancy into a
  conservation priorisation approach.
\newblock Master's thesis, University Of East Anglia, Norwich, UK.

\bibitem[Redding et~al., 2008]{red-har-08}
Redding, D.~W., Hartmann, K., Mimoto, A., Bokal, D., DeVos, M., and Mooers, A.
  (2008).
\newblock Evolutionarily distinctive species often capture more phylogenetic
  diversity than expected.
\newblock {\em Journal of Theoretical Biology}, 251(4):606--615.

\bibitem[Redding et~al., 2014]{red-maz-14}
Redding, D.~W., Mazel, F., and Mooers, A.~{\O}. (2014).
\newblock Measuring evolutionary isolation for conservation.
\newblock {\em {PLoS} {ONE}}, 9(12):e113490.

\bibitem[Redding and Mooers, 2006]{red-moo-06a}
Redding, D.~W. and Mooers, A.~{\O}. (2006).
\newblock Incorporating evolutionary measures into conservation prioritization.
\newblock {\em Conservation Biology}, 20(6):1670--1678.

\bibitem[Semple and Steel, 2003]{SS03}
Semple, C. and Steel, M. (2003).
\newblock {\em Phylogenetics}.
\newblock Oxford University Press.

\bibitem[Shapley, 1953]{sha-53a}
Shapley, L.~S. (1953).
\newblock A value for $n$-person games.
\newblock {\em Contributions to the Theory of Games}, 2(28):307--317.

\bibitem[Spillner et~al., 2008]{spi-ngu-08a}
Spillner, A., Nguyen, B.~T., and Moulton, V. (2008).
\newblock Computing phylogenetic diversity for split systems.
\newblock {\em IEEE/ACM Transactions on Computational Biology and
  Bioinformatics}, 5(2):235--244.

\bibitem[Steel, 2016]{ste-16}
Steel, M. (2016).
\newblock {\em Phylogeny: Discrete and random processes in evolution}.
\newblock SIAM, Philadelphia PA.

\bibitem[Tucker et~al., 2016]{tuc-cad-16}
Tucker, C.~M., Cadotte, M.~W., Carvalho, S.~B., Davies, T.~J., Ferrier, S.,
  Fritz, S.~A., Grenyer, R., Helmus, M.~R., Jin, L.~S., Mooers, A.~O., Pavoine,
  S., Purschke, O., Redding, D.~W., Rosauer, D.~F., Winter, M., and Mazel, F.
  (2016).
\newblock A guide to phylogenetic metrics for conservation, community ecology
  and macroecology.
\newblock {\em Biological Reviews}, 92(2):698--715.

\bibitem[Vane-Wright et~al., 1991]{van-wri-91}
Vane-Wright, R., Humphries, C., and Williams, P. (1991).
\newblock What to protect?{\textemdash}{S}ystematics and the agony of choice.
\newblock {\em Biological Conservation}, 55(3):235--254.

\bibitem[Volkmann et~al., 2014]{vol-mar-14a}
Volkmann, L., Martyn, I., Moulton, V., Spillner, A., and Mooers, A.~O. (2014).
\newblock Prioritizing populations for conservation using phylogenetic
  networks.
\newblock {\em PLoS One}, 9(2).

\bibitem[Wicke and Fischer, 2018]{wic-fis-18a}
Wicke, K. and Fischer, M. (2018).
\newblock Phylogenetic diversity and biodiversity indices on phylogenetic
  networks.
\newblock {\em Mathematical Biosciences}, 298:80--90.

\bibitem[Wicke and Steel, 2019]{wic-ste-19a}
Wicke, K. and Steel, M. (2019).
\newblock Combinatorial properties of phylogenetic diversity indices.
\newblock {\em Journal of Mathematical Biology}, pages 1--29.

\end{thebibliography}

\newpage

\section*{Appendix}

\begin{figure}[h!]
\centering
\includegraphics[scale=0.85]{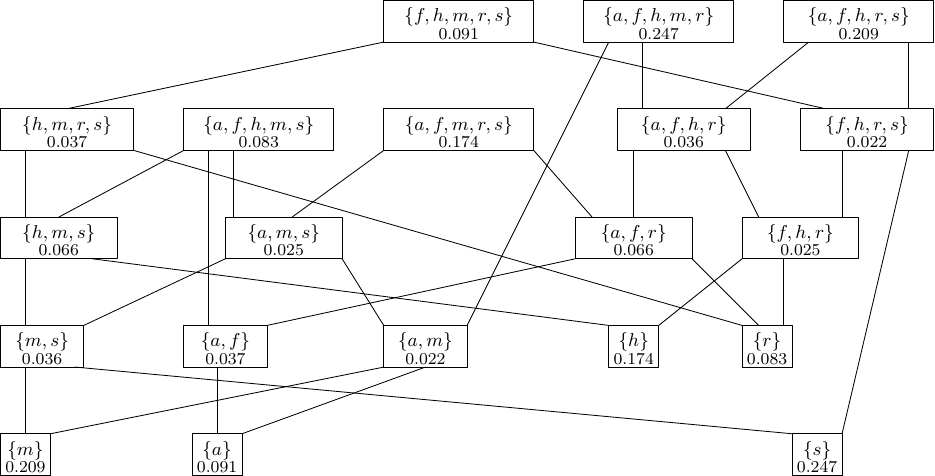}
\caption{The Hasse diagram for the clusters in 
the cluster system \(\mathcal{C}(\mathcal{S})\)
on~\(X = \{a,f,h,m,r,s\}\) computed
from the split system \(\mathcal{S}\) on~\(X\)
represented by the phylogenetic network in Figure~\ref{fig:nnet:owls}.
The number below each cluster is the weight of the cluster
obtained by the map~\(\tau\) from the weights of the splits
in~\(\mathcal{S}\).}
\label{fig:hasse:owls}
\end{figure}

\begin{figure}[h!]
\centering
\includegraphics[scale=0.85]{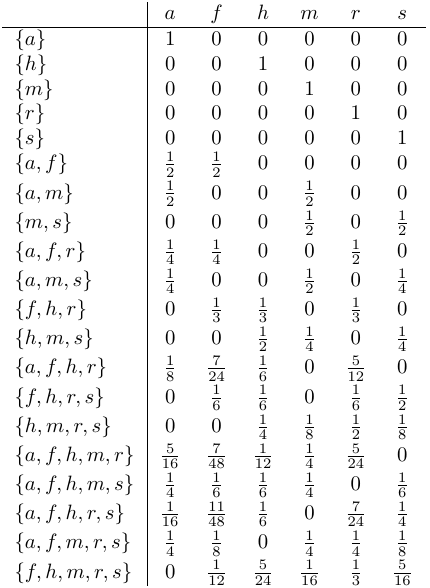}
\caption{The matrix \(\Gamma\) for the equal splits index
on \(\mathbb{L}(\mathcal{C})\) for the cluster system \(\mathcal{C}\) in
Figure~\ref{fig:hasse:owls}.}
\label{fig:gamma:es:owls}
\end{figure}

\end{document}